# The Nonlinear Redshift Space Power Spectrum: Ω from Redshift Surveys

Karl B. Fisher[1] & Adi Nusser[2]
[1] Institute for Advanced Study, Olden Lane, Princeton, NJ 08540
[2] Institute of Astronomy, Madingley Rd., Cambridge, CB3 0HA, England

**ABSTRACT**
We examine the anisotropies in the power spectrum by the mapping of real to redshift space. Using the Zel'dovich approximation, we obtain an analytic expression for the nonlinear redshift space power spectrum in the distant observer limit. For a given unbiased galaxy distribution in redshift space, the anisotropies in the power spectrum depend on the parameter $f(\Omega) \approx \Omega^{0.6}$, where $\Omega$ is the density parameter. We quantify these anisotropies by the ratio, $R$, of the quadrupole to monopole angular moments of the power spectrum. In contrast to linear theory, the Zel'dovich approximation predicts a decline in $R$ with decreasing scale. This departure from linear theory is due to nonlinear dynamics and not a result of incoherent random velocities. The rate of decline depends strongly on $\Omega$ and the initial power spectrum. However, we find a *universal* relation between the quantity $R/R_{lin}$ (where $R_{lin}$ the linear theory value of $R$) and the dimensionless variable $k/k_{nl}$, where $k_{nl}$ is a wavenumber determined by the scale of nonlinear structures. The universal relation is in good agreement with a large N-body simulation. This universal relation greatly extends the scales over which redshift distortions can be used as a probe of $\Omega$. A preliminary application to the 1.2 Jy IRAS yields $\Omega \sim 0.4$ if IRAS galaxies are unbiased.

## 1 INTRODUCTION

Statistical studies of the large scale structure are mainly based on galaxy redshift surveys. Redshifts of galaxies differ from their distances due to peculiar velocities along the line of sight. This difference produces anisotropies in the observed galaxy distribution in redshift-space (s-space). The anisotropies of clustering in s-space are manifested in the dependence of the correlation function on the direction of the pair separation or, equivalently, in the dependence of the power spectrum on the direction of the wavevector. These anisotropies offer the promise of measuring the cosmological density parameter, $\Omega$, on large scales. This is because gravitational instability predicts that the peculiar velocity causing the large scale distortions in s-space depend on $f(\Omega) \equiv d\ln D/d\ln a \approx \Omega^{0.6}$, where $D(t)$ and $a(t)$ are the linear growth and scale factors respectively. Therefore, given an approximation for gravity, one may obtain an estimate of $\Omega$ by quantifying the anisotropies, say, in the observed s-space power spectrum.

A useful way to characterize the anisotropies in s-space is through the angular moments of the s-space power spectrum, $P^s$; the lowest order moments, the monopole and quadrupole, are given by

$$M = \int\limits_0^{+1} d\mu \; P^s(k,\mu)$$

and

$$Q = \frac{5}{2} \int\limits_0^{+1} d\mu \; P^s(k,\mu) \, (3\mu^2 - 1), \quad (1)$$

where $\mu$ is the angle between the wavevector **k** and the line of sight. In the linear regime, the ratio $R \equiv Q/M$ is independent of scale and given by

$$R_{lin} = \frac{\frac{4}{3}\beta + \frac{4}{7}\beta^2}{1 + \frac{2}{3}\beta + \frac{1}{5}\beta^2}, \quad (2)$$

where $\beta = f(\Omega)/b$, and $b$ is the linear bias factor. Values of $\beta$ determined from s-space anisotropies span the range $\sim 0.4 - 1.0$ (Cole, Fisher, & Weinberg 1995 (CFW); Hamilton 1995; Fisher, Scharf, & Lahav 1994; Heavens & Taylor 1995; Fisher et al. 1994; Loveday et al. 1995). One source of systematic error in these analyses is in modeling the nonlinear gravitational evolution (Cole, Fisher, & Weinberg 1994). In this *Letter*, we study the nonlinear evolution of the s-space power spectrum. To describe nonlinear dynamics we adopt the Zel'dovich approximation (Zel'dovich 1970). This approximation is an excellent tool for describing nonlinear dynamics on large scales (*e.g.*, Nusser *et al.* 1991). The Zel'dovich approximation has previously been used by Taylor (1993) and Schneider & Bartelmann (1995) to study the evolution of the real space power spectrum.

In §2 we derive an analytic expression for the s-space power spectrum and find a strong dependence of the ratio $R$ on $\Omega$ and the initial conditions. However, in §3, we show that the quantity $R/R_{lin}$ when plotted against the dimensionless variable $k/k_{nl}$, where $k_{nl}$ is a wavenumber determined by the scale of nonlinear structures, acquires a universal shape independent of $\Omega$ and the initial power spectrum. This relation can be very useful for estimating $\Omega$ from redshift surveys over a wide range of scales. As an example, we present a preliminary application to the 1.2 Jy IRAS survey. We conclude in §4.

## 2 THE REDSHIFT SPACE POWER SPECTRUM IN THE ZEL'DOVICH APPROXIMATION

Let $\mathbf{v}$, $\mathbf{q}$, $\mathbf{x}$ and $\mathbf{s}$ be, respectively, the comoving peculiar velocity, the initial coordinate (*i.e*, Lagrangian), the present real-space (r-space) coordinate (*i.e*, Eulerian), and the s-space coordinate of a particle in units of km s$^{-1}$. The Zel'dovich approximation states that the displacement vector, $\mathbf{d}$, from $\mathbf{q}$ to $\mathbf{x}$ is related to the peculiar velocity by $f(\Omega)^{-1}\mathbf{v}$; accordingly the s-space coordinate, $\mathbf{s}$, is given by

$$\mathbf{s} = \mathbf{x} + (\mathbf{v}\cdot\hat{\mathbf{l}})\hat{\mathbf{l}} = \mathbf{q} + (\mathbf{I} + f(\Omega)\,\mathcal{P})\,\mathbf{d}(\mathbf{q},t) \quad , \quad (3)$$

where $\mathbf{I}$ is the unity matrix and $\mathcal{P} = \hat{\mathbf{l}}\hat{\mathbf{l}}^{\mathrm{T}}$ is a matrix which projects the velocity vector along the line of sight denoted by the unit vector $\hat{\mathbf{l}}$. In order to simplify the calculations, we work in the "distant-observer" limit where the line of sight is well approximated by a fixed direction. Having specified the mapping (3), we can readily write down a formal equation for the matter density field in s-space as,

$$\delta^s(\mathbf{s}) = \int d^3\mathbf{q}\,\delta_D\left[\mathbf{s} - \mathbf{q} - (\mathbf{I} + f\,\mathcal{P})\mathbf{d}\right] - 1 \quad , \quad (4)$$

where the $s$ superscript denotes s-space and $\delta_D$ is the Dirac delta function.

By Fourier transforming (4) we find,

$$\begin{aligned}\delta^s_{\mathbf{k}} &= \int d^3\mathbf{s}\,\exp[i\mathbf{k}\cdot\mathbf{s}]\,\delta^s(\mathbf{s}) \\ &= \int d^3\mathbf{q}\,\exp[i\mathbf{k}\cdot\mathbf{q}]\,\left\{\exp[i\mathbf{k}^{\mathrm{T}}(\mathbf{I} + f\,\mathcal{P})\mathbf{d}] - 1\right\}\end{aligned} \quad (5)$$

The second term in brackets reflects the Fourier transform of the mean density and contributes only to the $\mathbf{k} = 0$ mode. From (5) we see that for $\mathbf{k} \neq 0$

$$\begin{aligned}\langle\delta^s_{\mathbf{k}}\delta^s_{\mathbf{k}'}\rangle &= \int d^3\mathbf{q}\,d^3\mathbf{q}'\,\exp[i\mathbf{k}\cdot\mathbf{q} + i\mathbf{k}'\cdot\mathbf{q}'] \\ &\quad\times\left\langle\exp\left[i\mathbf{k}^{\mathrm{T}}(\mathbf{I} + f\,\mathcal{P})\mathbf{d} + i\mathbf{k}'^{\mathrm{T}}(\mathbf{I} + f\,\mathcal{P})\mathbf{d}'\right]\right\rangle\end{aligned} \quad , \quad (6)$$

where $\mathbf{d}' \equiv \mathbf{d}(\mathbf{q}')$. The covariance of the modes is related to the s-space power spectrum, $P^s$, by

$$\langle\delta^s_{\mathbf{k}}\delta^s_{\mathbf{k}'}\rangle \equiv (2\pi)^3\delta_D(\mathbf{k} + \mathbf{k}')\,P^s(\mathbf{k}) \quad , \quad (7)$$

where the Dirac delta is a result of translational invariance insured by working in the distant observer limit. The relations (6) and (7) specify the s-space power spectrum in the Zel'dovich approximation.

The ensemble average in (6) can be written in terms of an integral over the joint probability distribution, $\mathrm{Pr}(\mathbf{d}, \mathbf{d}')$, of the displacement vectors

$$\begin{aligned}&\int d^3\mathbf{d}\,d^3\mathbf{d}'\,\mathrm{Pr}(\mathbf{d},\mathbf{d}') \\ &\quad\times\exp\left[i\mathbf{k}^{\mathrm{T}}(\mathbf{I} + f\mathcal{P})\mathbf{d} + i\mathbf{k}'^{\mathrm{T}}(\mathbf{I} + f\mathcal{P})\mathbf{d}'\right].\end{aligned} \quad (8)$$

The great simplification of the Zel'dovich approximation is that the displacement is proportional to the initial peculiar velocity. If we assume that the initial fluctuations were Gaussian then, $\mathrm{Pr}(\mathbf{d}, \mathbf{d}')$ is specified by the correlation function of the displacement field, $\mathbf{C}(\mathbf{r}) = \langle\mathbf{d}\mathbf{d}'^T\rangle$, where $\mathbf{r} = \mathbf{q}' - \mathbf{q}$. After performing the integral in (8), and comparing with (7) we find

$$\begin{aligned}P^s(\mathbf{k}) &= \int d^3\mathbf{r}\,e^{-i\mathbf{k}\cdot\mathbf{r}} \\ &\quad\times\exp[\mathbf{k}^{\mathrm{T}}(\mathbf{I} + f\,\mathcal{P})(\mathbf{C}(\mathbf{r}) - \mathbf{C}(0))(\mathbf{I} + f\,\mathcal{P})\mathbf{k}] \quad .\end{aligned} \quad (9)$$

The displacement field correlation function can be conveniently written in in following form (e.g., Górski 1988)

$$C_{ij}(\mathbf{r},t) = \delta_{ij}C_{\perp}(r,t) + \hat{r}_i\hat{r}_j\left[C_{\parallel}(r,t) - C_{\perp}(r,t)\right] \quad , \quad (10)$$

where

$$C_{\perp}(r,t) = \frac{D(t)^2}{2\pi^2}\int dk\,P_{lin}(k)\frac{j_1(kr)}{kr} \quad (11)$$

$$C_{\parallel}(r,t) = \frac{D(t)^2}{2\pi^2}\int dk\,P_{lin}(k)\left[j_0(kr) - \frac{2j_1(kr)}{kr}\right] \quad (12)$$

where $P_{lin}(k)$ is the initial r-space power spectrum.

To first order in $\mathbf{C}$, equation (9) reduces to

$$P^s_{lin}(k,\mu) = D(t)^2 P_{lin}(k)\,(1 + f\,\mu^2)^2 \quad . \quad (13)$$

This is the well known result of linear theory (Kaiser 1987); the agreement is not surprising since to lowest order the Zel'dovich approximation is equivalent to linear theory.

So far we have assumed that galaxies trace the matter distribution. In reality, we should allow for some bias between the galaxy and matter distributions. Here we make the hypothesis that when the fluctuations are small the galaxy and matter density fluctuations are related by a linear biasing relation. However, with the growth of fluctuations into the nonlinear regime at the present time, the bias factor becomes a nonlinear function of the density. With this hypothesis, the parameter $f(\Omega)$ in (9) is simply replaced by $\beta_{lin} = f(\Omega)/b_{lin}$, where $b_{lin}$ is the initial linear bias factor scaled to the present using linear theory.

The integral in (9) can be numerically evaluated to yield the nonlinear s-space power spectrum for a given initial power spectrum and a value of $\Omega$. We introduce a sharp small scale cutoff, $k_c$, in the initial power spectrum in an attempt to remedy the failure of the Zel'dovich approximation in regions where orbit mixing has occured. We consider a family of CDM-like models specified by the shape parameter $\Gamma$ as defined in Efstathiou, Bond, & White (1992). We quantify the anisotropies in $P^s$ at each $|\mathbf{k}|$ by the ratio $R = Q/M$ where $Q$ and $M$ are computed according to (1). Figure 1 shows $R$ for a variety of different $\Gamma$, $\sigma_8$, and $\Omega$ (cf. caption). All curves but the light solid one were generated using a small scale cutoff of $2\pi/k_c = 5\ h^{-1}$Mpc. For comparison, the light solid curve was derived with a smaller truncation of $2\pi/k_c = 1\ h^{-1}$Mpc. We see that, fortunately, $R$ is only weakly sensitive to the cutoff.

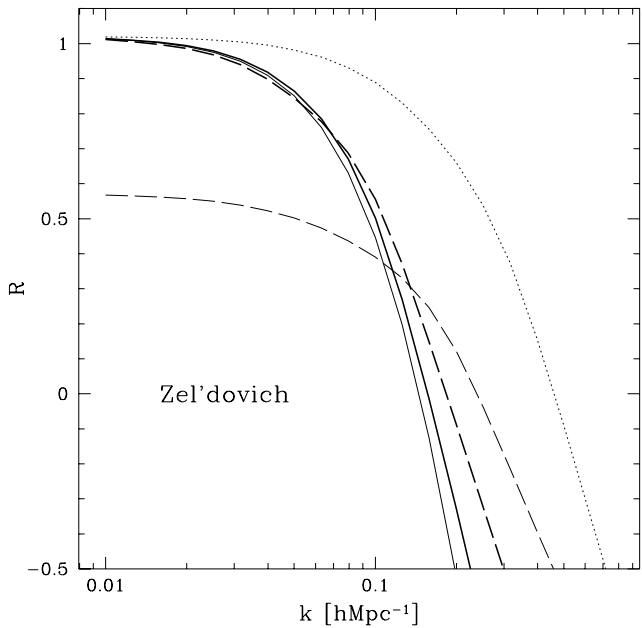

**Figure 1.** The ratio, $R = Q/M$, computed from the Zel'dovich approximation. The heavy solid and dotted curves correspond to an initial power spectrum of ($\Gamma = 0.5, \Omega = 1$) with $\sigma_8 = 1$ and 0.5 respectively. The heavy and light dashed curves correspond to ($\Gamma = 0.2, \sigma_8 = 1.0$) with $\Omega = 1$ and 0.3 respectively. The truncation of the initial power spectra of these curves is $2\pi/k_c = 5\ h^{-1}$Mpc. The light solid curve is the same as the heavy solid curve but with truncation at $2\pi/k_c = 1\ h^{-1}$Mpc.

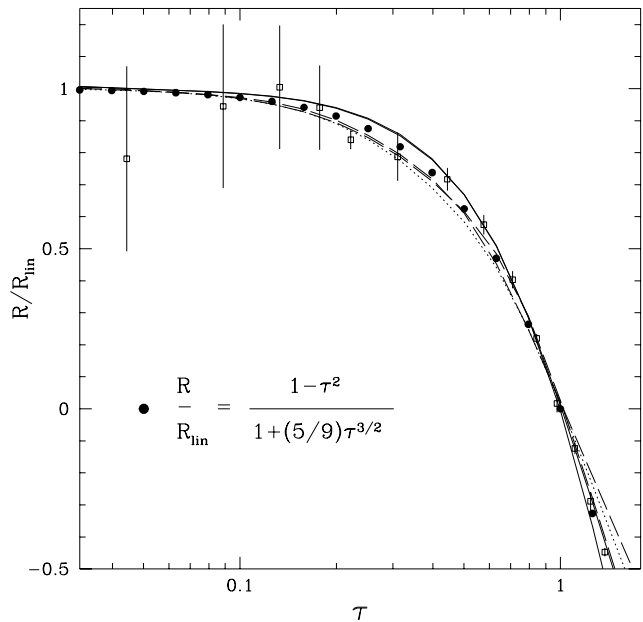

**Figure 2.** The ratio, $R/R_{lin}$, from the curves in figure 1 plotted against the dimensionless variable $\tau = k/k_{nl}$. The solid dots show the empirical fit (14) to the universal relation. The open squares with errorbars were obtained from the moments given in CFW for their 400 $h^{-1}$Mpc, $\Gamma = 0.25$, $\sigma_8 = 0.8$, and $\Omega = 0.3$ N-body simulation (cf., CFW figure 2).

On large scales, $R$ asymptotes to $R_{lin}$ given in (2), but decreases rapidly with increasing wavenumber. This decline is because linear theory over predicts the peculiar velocity associated with a given density field (Nusser *et al.* 1991) and consequently overestimates the amplitude of anisotropies in the power spectrum on scales where nonlinear effects are important. Our formalism accounts for multi-valued zones in which the flow in r-space is laminar even though particle orbits have crossed in s-space. On scales where this effect is significant, the quadrupole moment reverses sign and $R$ becomes negative. It is important to distinguish this effect from that of incoherent velocities in virialised regions (Fingers of God). The scale at which $R$ crosses zero defines a nonlinear scale corresponding to a wavenumber $k_{nl}$. The value of $k_{nl}$ is determined by the initial power spectrum as well as $\Omega$. The $\Omega$ dependence is due to the fact that for a given r-space density field the predicted velocity is proportional to $f(\Omega)$. Hence the nonlinear scale in s-space is a function of $\Omega$.

## 3 UNIVERSAL SCALING RELATION

In the Zel'dovich approximation the peculiar velocity is specified by the r-space density field independent of the initial conditions (Nusser *et al.* 1991). Since the anisotropies in s-space are a reflection of this density/velocity relation, we expect the ratio $R$ to be a weak function of the initial power spectrum when expressed in terms of the dimensionless variable $\tau \equiv k/k_{nl}$. In order to verify this ansatz, we plot, in figure 2, the ratio $R/R_{lin}$ against the variable $\tau$ for each of the curves in figure 1. In addition, we show results from the large N-body simulation of CFW. This figure shows that, indeed, the ratio $R/R_{lin}$ acquires a universal shape over a large range of scales (over two orders of magnitude). The agreement of the Zel'dovich predictions with the simulation is remarkable. The small differences between the different curves may be attributed to two effects. First, the nonlinear density/velocity relation in Fourier space involves coupling between different modes; this coupling may depend on the initial power spectrum. Second, the failure of Zel'dovich approximation in regions where orbit mixing has occured is especially pronounced for models with large power on small scales; in reality, however, caustics are preserved and therefore one might expect the scaling relation to hold more tightly.

A simple empirical fit to the scaling relation, shown by the solid dots in figure 2, is

$$\frac{R}{R_{lin}} = \frac{1 - \tau^2}{1 + \frac{5}{9}\tau^{3/2}} \quad . \tag{14}$$

This relation permits an estimation of the parameter $\Omega$ or more precisely $\beta_{lin}$ on scales where nonlinear effects are important. Here we use the relation (14) to estimate $\beta_{lin}$ from the 1.2 Jy IRAS survey (Fisher *et al.* 1995). The ratio $R$ for this survey versus scale was taken from CFW (figure 7). The wavenumber at which $R$ crosses zero was estimated to be $k_{nl} = 0.22 \, h\mathrm{Mpc}^{-1}$. The three solid curves in the figure show the value of $R$ determined from the scaling relation (14) for different $\beta_{lin}$. Though the errors in $R$ are large, the data favor $\beta_{lin} \sim 0.6$.

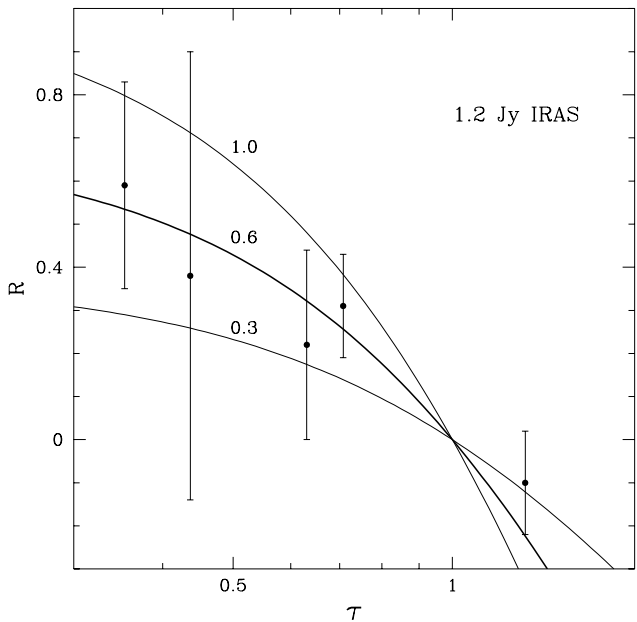

**Figure 3.** The ratio, $R$, from the 1.2 Jy IRAS survey plotted against the variable $\tau$. The points were obtained from figure 7 in CFW. The curves correspond to the universal scaling model (equation 14) with values of $\beta_{lin} = 0.3$, 0.6, and 1.0 as denoted on the figure.

## 4  DISCUSSION

We have presented a calculation of the anisotropies in the s-space power spectrum based on the Zel'dovich approximation. The Zel'dovich approximation shows clearly that the decline of the quadrupole to monopole ratio on large scales is due to nonlinear, yet coherent, peculiar velocities. Despite a strong dependence on $\Omega$ and the initial power spectra, the s-space anisotropies obey a universal scaling relation when expressed as a function of the variable $k/k_{nl}$. This scaling relation extends analyses of redshift distortions into the nonlinear regime with the only free parameter being $\beta_{lin}$. This is a marked improvement over phenomenological models for the distortion which require an unrealistically high random velocity component to explain the decline of $R$ with scale (*e.g.*, Peacock & Dodds 1994; CFW).

We have not incorporated the effects of random velocities. The main manifestation of random velocities is the so-called Finger of God effect arising from the redshift stretching of virialised regions of rich clusters. The velocity dispersion in these regions can be on the order of $\sim 1000$ km s$^{-1}$. Yet, rich clusters are rare in the universe and can, in principle, be excised from the analysis. Field galaxies have a random velocity component in addition to their coherent motions. However, the amplitude of this component is small. The one-dimensional velocity dispersion of all galaxies, including those in clusters, is estimated to be $\approx 140$–$200$ km s$^{-1}$ (Miller, Davis & White 1995).

Measurements of the redshift distortions in the planned Sloan Digital Sky Survey (Gunn & Knapp 1993) and Anglo-Australian 2dF galaxy survey should provide accurate determinations of $R$ over a wide range of scales. Our formalism is especially suitable for the estimation of $\Omega$ from these surveys.

## ACKNOWLEDGMENTS

KBF gratefully acknowledges the financial support of the Ambrose Monell Foundation. AN acknowledges the support of a PPARC research fellowship. We thank Shaun Cole and David Weinberg for providing results from the N-body simulation.